\documentclass[preprint]{elsarticle}
\usepackage{lineno,hyperref}
\usepackage{amsmath}
\usepackage{amssymb}
\usepackage{setspace}
\usepackage{epsfig}
\usepackage{epstopdf}
\usepackage{rotating}
\usepackage{caption,subcaption}
\usepackage{color}
\usepackage{colortbl}
\usepackage{graphicx}
\usepackage{ifthen}
\usepackage{alltt}
\usepackage{float}
\usepackage{fancyhdr}
\usepackage{verbatim}
\usepackage{moreverb}
\usepackage{multirow}
\usepackage{bigstrut}
\usepackage{subcaption}

\modulolinenumbers[5]

\def\_#1{{\bf #1}}

\def\.{\cdot}
\def\x{\times}

\def\l#1{\label{eq:#1}}
\def\r#1{(\ref{eq:#1})}

\def\e{\begin{equation}}
\def\f{\end{equation}}
\def\px{ p_1}
\def\py{p_2}
\def\mx{m_1}
\def\my{m_2}
\def\Ex{E_{\rm inc 1}}
\def\Ey{E_{\rm inc 2}}
\def\Hx{H_{\rm inc 1}}
\def\Hy{H_{\rm inc 2}}
\def\eexx{\alpha^{\rm 11}_{\rm ee}}
\def\eexy{\alpha^{\rm 12}_{\rm ee}}
\def\eeyx{\alpha^{\rm 21}_{\rm ee}}
\def\eeyy{\alpha^{\rm 22}_{\rm ee}}
\def\emxx{\alpha^{\rm 11}_{\rm em}}
\def\emxy{\alpha^{\rm 12}_{\rm em}}
\def\emyx{\alpha^{\rm 21}_{\rm em}}
\def\emyy{\alpha^{\rm 22}_{\rm em}}
\def\mexx{\alpha^{\rm 11}_{\rm me}}
\def\mexy{\alpha^{\rm 12}_{\rm me}}
\def\meyx{\alpha^{\rm 21}_{\rm me}}
\def\meyy{\alpha^{\rm 22}_{\rm me}}
\def\mmxx{\alpha^{\rm 11}_{\rm mm}}
\def\mmxy{\alpha^{\rm 12}_{\rm mm}}
\def\mmyx{\alpha^{\rm 21}_{\rm mm}}
\def\mmyy{\alpha^{\rm 22}_{\rm mm}}
\def\H0{{H_0}}
\def\E0{\eta {H_0}}
\def\n{\eta}

\def\=#1{\overline{\overline #1}}

\journal{Photonics and Nanostructures}

\begin{document}

\begin{frontmatter}

\title{Determining polarizability tensors for an arbitrary small electromagnetic scatterer}

\author[Finland,Belarus]{Viktar S.~Asadchy}
\cortext[mycorrespondingauthor]{Corresponding author}
\ead{viktar.asadchy@aalto.fi}
\author[Belarus]{Igar A.~Faniayeu}
\author[Finland]{Younes Ra'di}
\author[Finland]{Sergei A.~Tretyakov}

\address[Finland]{Department of Radio Science and Engineering, Aalto University, FI-00076 Aalto, Finland}
\address[Belarus]{Department of General Physics, Gomel State University, Sovyetskaya Str.~104, 246019, Gomel, Belarus}

\begin{abstract}
In this paper, we present a method to retrieve tensor polarizabilities of general bi-anisotropic particles from their far-field responses to plane-wave illuminations. The necessary number of probing excitations and the directions where the scattered fields need to be calculated or measured have been found. When implemented numerically, the method does not require any spherical harmonic expansion nor direct calculation of dipole moments, but only calculations of co- and cross-polarized scattering cross sections for a number of plane-wave excitations. With this simple approach, the polarizabilities can be found also from experimentally measured cross sections. The method is exemplified considering two bi-anisotropic particles, a reciprocal omega particle and a non-reciprocal particle containing a ferrite inclusion coupled to metal strips. 
\end{abstract}


\end{frontmatter}


\section{Introduction}
Artificial materials (metamaterials) made of small inclusions (meta-atoms) positioned beside each other have become very popular, because with this concept it is possible to realize exotic electromagnetic properties which are not found in natural materials. These inclusions, so-called meta-atoms, which are electrically small (in comparison to the wavelength in the medium), in the sense of their electromagnetic response play the same role as atoms do in natural materials. The averaged electric and magnetic properties of a metamaterial sample are determined by electric and magnetic properties of individual inclusions and by their mutual interactions. The meta-atoms can be characterized by their polarizabilities. The polarizabilities show how a single meta-atom behaves in responding to external electromagnetic fields. Knowing the electromagnetic properties of each building block of the metamaterial allows us to understand the electromagnetic properties of metamaterials as composite media. In particular, proper engineering of meta-atoms allows us to design metamaterials with required effective properties. There are several approaches to determination of effective parameters of a medium knowing the electric and magnetic moments of each building block of the medium, both for volumetric (bulk) samples and thin layers (metasurfaces), e.g. \cite{Lewin,Sihvola,analytical,Holloway,Alu_ho,Teemu}. In this paper, we discuss how the meta-atom polarizabilities can be retrieved from the knowledge of single meta-atom far-field scattering response to plane-wave excitations. 

The spherical harmonic expansion theory introduced by Mie for a homogeneous sphere of any size and arbitrary refractive index is known as one of the main tools in deriving the polarizabilities of a single dielectric sphere \cite{Mie,Lorenx,Bohren}. Later, this theory was extended to particles of an arbitrary shape \cite{Waterman}. Recently, the extended theory has been used by many researchers to study the multipolar behavior of special inclusions, e.g. \cite{Menzel,Arango}. Using the generalized Mie theory and writing the scattered fields in terms of vector spherical harmonics, multipolar moments of an arbitrary scatterer can be calculated. However, this approach implies computationally heavy integrations of scattered fields over the sphere surrounding the particle that complicates the implementation of the method in numerical calculations. Furthermore, it appears problematic to use such methods for extracting polarizabilities from experimentally measured response of the particle.

In most cases when the particle is electrically small, electric and magnetic dipolar moments are the only significant and important moments in the Mie expansion. This assumption allows us to dramatically simplify the scattering-based polarizability retrieval and propose a much simpler method for extracting polarizability tensors of an arbitrary small scatterer from its response in the far zone. To extract one specific polarizability component of the scatterer, our method implies determination of the scattered fields only in two special directions. This significantly simplifies the realization of the method in numerical calculations. Furthermore, the discrete and minimal number of directions in which the scattered fields must be probed allows us to utilize the method also experimentally. This method for the first time was proposed in \cite{Fanyaev} for helical particles possessing bi-anisotropic electromagnetic coupling. In the present paper, we generalize the polarizability retrieval method so that it can be utilized for arbitrary small particles with the most general bi-anisotropic properties. 
The method can be considered as a generalization of the approach used in \cite{antenna} for determination of the polarizablities of small chiral particles from their co- and cross-polarized scattering cross sections.

In the most general case, assuming that the induced dipole moments in the particle depend linearly on the  applied fields, the dipolar moments induced in the particle relate to the incident fields (at the location of the particle) by the polarizability tensors as:
\e
\begin{array}{c}\displaystyle
\mathbf{p}=\overline{\overline{\alpha}}_{\rm ee}\.\mathbf{E}_{\rm inc}+\overline{\overline{\alpha}}_{\rm em}\.\mathbf{H}_{\rm inc},\vspace{.1cm}\\\displaystyle
\mathbf{m}=\overline{\overline{\alpha}}_{\rm me}\.\mathbf{E}_{\rm inc}+\overline{\overline{\alpha}}_{\rm mm}\.\mathbf{H}_{\rm inc}.
\end{array}\l{moments}\f
These relations hold for bi-anisotropic particles of all known classes: reciprocal chiral and omega, non-reciprocal Tellegen and ``moving'' particles, and any combination of these \cite{serdukov}, \cite{tretyakov32}. For a special case of anisotropic particles without electromagnetic coupling (e.g., small dielectric spheres) the relations are simplified taking into account that $\overline{\overline{\alpha}}_{\rm em}=\overline{\overline{\alpha}}_{\rm me}=0$.

The structure of the paper is as follows. In Section~\ref{sec:2}, we formulate the basic idea and derive the proper expressions for the polarizabilities of a general bi-anisotropic particle (assuming the particle is electrically small). In Section~\ref{sec:3}, we implement the method for two different particles: a reciprocal omega particle and a non-reciprocal particle possessing moving and chiral electromagnetic couplings.

\section{Basic formulation}\label{sec:2}

\subsection{Polarizabilities of a bi-anisotropic particle}\label{sec:2a}

\begin{figure}[h]
\centering
\epsfig{file=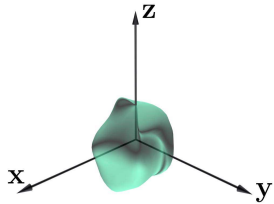, width=0.3\textwidth}
\caption{An arbitrary particle in the center of a Cartesian coordinate system.}
\label{fig:particle}
\end{figure}

In order to determine the polarizabilities of an arbitrary particle, we analyze the far-field response of the particle to incident plane waves. We start from writing the relations for the polarizabilities of the particle in terms of the dipole moments induced by a set of probing fields. Let us fix the position of the particle at the center of a Cartesian coordinate system (see Fig.~\ref{fig:particle}).
The particle is situated in free space with the characteristic impedance $\eta_0$.
As an example, here we write the formulas for the case when the incident plane waves illuminating the particle propagate along the $\_z_0$-axis. It will be shown that the directions can be chosen arbitrarily. Obviously, choosing the $\_z_0$-directed incident waves, one can find only the components of the polarizability tensors in the $\_x_0\_y_0$-plane. The other components can be determined similarly using incident waves propagating along the $\_x_0$ and $\_y_0$ axes.
Taking into account that the incident plane waves are transverse and propagate along the $\_z_0$-axis, equations \r{moments} simplify to:
\e\begin{array}{c}\displaystyle
\left[ \begin{array}{c} \px \\ \py \end{array} \right]
=\left[ \begin{array}{cc}
\eexx &\eexy \\ \eeyx & \eeyy \end{array} \right]\cdot
\left[ \begin{array}{c} \Ex \\ \Ey \end{array} \right] +
\left[ \begin{array}{cc}
\emxx &\emxy \\ \emyx & \emyy \end{array} \right]\cdot
\left[ \begin{array}{c} \Hx \\ \Hy \end{array} \right],
\vspace{.5cm}\\\displaystyle

\left[ \begin{array}{c} \mx \\ \my \end{array} \right]
=\left[ \begin{array}{cc}
\mexx &\mexy \\ \meyx & \meyy \end{array} \right]\cdot
\left[ \begin{array}{c} \Ex \\ \Ey \end{array} \right] +
\left[ \begin{array}{cc}
\mmxx &\mmxy \\ \mmyx & \mmyy \end{array} \right]\cdot
\left[ \begin{array}{c} \Hx \\ \Hy \end{array} \right].
\end{array}\l{eq:a}\f
Hereafter we use numerical indices $1, 2, 3,$ representing the $\_x_0$, $\_y_0$ and $\_z_0$ projections, respectively.
It can be seen from \r{eq:a} that to determine any polarizability component, it is insufficient to know the response of the particle to only one incident wave. The simplest way to find the component is illumination of the particle by two incident plane waves with the following polarization states:
\e
\begin{array}{c}
\mathbf{E}_{\rm inc}=\eta_0 {H_0} \mathbf{x_0},\qquad
\mathbf{H}_{\rm inc}=\pm {H_0} \mathbf{y_0},
\end{array}\l{eq:b}\f
in which the upper and lower signs correspond to waves propagating in $+\_z_0$ and  $-\_z_0$ directions, respectively, and $H_0$ is the magnitude of the incident magnetic field. Here, for simplicity, we assume that the two incident waves have equal amplitudes and phases at the location of the particle. Although in practice it is difficult to generate two incident waves with precisely equal phases, the problem can be solved similarly with the assumption that the waves in \r{eq:b} have not only different propagation directions, but also different amplitudes and phases, meaning that this assumption is not limiting.
Substituting \r{eq:b} in \r{eq:a}, we get 8 equations:
\e
\begin{array}{c}\displaystyle
p^{\pm}_1=\eexx \E0\pm\emxy \H0,\qquad \displaystyle
p^{\pm}_2=\eeyx \E0\pm\emyy \H0,\vspace{.3cm}\\\displaystyle
m^{\pm}_1=\mexx \E0\pm\mmxy \H0,\qquad\displaystyle
m^{\pm}_2=\meyx \E0\pm\mmyy \H0,
\end{array}\l{eq:c}\f
where the double signs correspond to the double signs in \r{eq:b}.
Next, the simple solution of the equations with regard to the polarizability components reads:
\e
\begin{array}{c}\displaystyle
\eexx=\frac{1}{2\E0} (p^{+}_{\rm 1}+p^{-}_{\rm 1}),\qquad \displaystyle
\emxy=\frac{1}{2\H0} (p^{+}_{\rm 1}-p^{-}_{\rm 1}),\vspace{.3cm}\\\displaystyle
\eeyx=\frac{1}{2\E0} (p^{+}_{\rm 2}+p^{-}_{\rm 2}),\qquad\displaystyle
\emyy=\frac{1}{2\H0} (p^{+}_{\rm 2}- p^{-}_{\rm 2}),\vspace{.3cm}\\\displaystyle

\mexx=\frac{1}{2\E0} (m^{+}_{\rm 1}+ m^{-}_{\rm 1}),\qquad \displaystyle
\mmxy=\frac{1}{2\H0} (m^{+}_{\rm 1}-m^{-}_{\rm 1}),\vspace{.3cm}\\\displaystyle
\meyx=\frac{1}{2\E0} (m^{+}_{\rm 2}+m^{-}_{\rm 2}),\qquad\displaystyle
\mmyy=\frac{1}{2\H0} (m^{+}_{\rm 2}-m^{-}_{\rm 2}).
\end{array}\l{eq:d}\f
In order to derive the other 8 polarizability components in the $\_x_0\_y_0$-plane, we choose the incidence in the form:
\e
\begin{array}{c}
\mathbf{E}_{\rm inc}=\eta_0{H_0} \mathbf{y_0},\qquad
\mathbf{H}_{\rm inc}=\pm {H_0} \mathbf{x_0}.
\end{array}\l{eq:e}\f
Likewise, for these two different incident waves, we can write 8 equations for the polarizabilities according to \r{eq:a}:
\e
\begin{array}{c}\displaystyle
\bar{p}^{\pm}_{\rm 1}=\eexy \E0\pm\emxx \H0,\qquad \displaystyle
\bar{p}^{\pm}_{\rm 2}=\eeyy \E0\pm\emyx \H0,\vspace{.3cm}\\\displaystyle
\bar{m}^{\pm}_{\rm 1}=\mexy \E0\pm\mmxx \H0,\qquad\displaystyle
\bar{m}^{\pm}_{\rm 2}=\meyy \E0\pm\mmyx \H0,
\end{array}\l{eq:f}\f
where we use notations with bars in order to distinguish the induced dipole moments for different polarization states in \r{eq:b} and \r{eq:e}. The double sign in \r{eq:f} corresponds to the double sign in \r{eq:e}.
Similarly, we can derive expressions for the polarizability components:
\e
\begin{array}{c}\displaystyle
\eexy=\frac{1}{2\E0} (\bar{p}^{+}_{\rm 1}+\bar{p}^{-}_{\rm 1}),\qquad \displaystyle
\emxx=\frac{1}{2\H0} (\bar{p}^{+}_{\rm 1}-\bar{p}^{-}_{\rm 1}),\vspace{.3cm}\\\displaystyle
\eeyy=\frac{1}{2\E0} (\bar{p}^{+}_{\rm 2}+\bar{p}^{-}_{\rm 2}),\qquad\displaystyle
\emyx=\frac{1}{2\H0} (\bar{p}^{+}_{\rm 2}-\bar{p}^{-}_{\rm 2}),\vspace{.3cm}\\\displaystyle

\mexy=\frac{1}{2\E0} (\bar{m}^{+}_{\rm 1}+\bar{m}^{-}_{\rm 1}),\qquad \displaystyle
\mmxx=\frac{1}{2\H0} (\bar{m}^{+}_{\rm 1}-\bar{m}^{-}_{\rm 1}),\vspace{.3cm}\\\displaystyle
\meyy=\frac{1}{2\E0} (\bar{m}^{+}_{\rm 2}+\bar{m}^{-}_{\rm 2}),\qquad\displaystyle
\mmyx=\frac{1}{2\H0} (\bar{m}^{+}_{\rm 2}-\bar{m}^{-}_{\rm 2}).
\end{array}\l{eq:g}\f
Thus, we have determined 16 polarizability components of the particle in terms of the induced dipole moments by probing plane waves. The other 20 components one can derive in the same way illuminating the particle by waves propagating along the $\_x_0$ and $\_y_0$ axes. In the next section, we determine the induced dipole moments in the particle from the far-zone scattered fields.

\subsection{Induced electric and magnetic dipole moments}\label{sec:2b}

A scatterer with induced oscillating electric and magnetic multipoles radiates energy in all directions.
Here, we study the case of an electrically small particle (the size of the particle is small compared to the wavelength of the incident waves) that allows us to take into account only the lowest multipoles, i.e. the electric and magnetic dipoles.
The scattered far fields from an electrically small particle are defined by the induced dipole moments in the form \cite{Definition}:
\e \begin{array}{c}\displaystyle
\mathbf{E}_{\rm sc}=\frac{k^2}{4\pi \epsilon_0 r} e^{-j k r}
\left[ (\mathbf{n}\x \mathbf{p})\x\mathbf{n} - \frac{1}{c \mu_0} \mathbf{n} \x \mathbf{m}\right],
\vspace{.1cm}\\\displaystyle
\mathbf{H}_{\rm sc}=\frac{1}{\eta_0}\mathbf{n} \x \mathbf{E}_{\rm sc},
\end{array}\l{eq:h}\f
where $\mathbf{n}$ is the unit vector in the direction of observation, $r$ is the distance between the particle and the observation point, $k=\omega / c$ is the wave number in surrounding space and the time-dependence $e^{j\omega t}$ is understood.
Since it is required to find only the $\_x_0$ and $\_y_0$ projections of the electric and magnetic dipole moments (according to \r{eq:d} and \r{eq:g}), we choose the observation direction to be along $+\_z_0$ and $-\_z_0$ (however, this choice is not compulsory). Taking this into account, we can rewrite the scattered electric field in \r{eq:h} as:
\e \begin{array}{c}\displaystyle
\mathstrut_{\rm z}\mathbf{E}_{\rm sc}=\gamma \hspace{1mm}
\left[ (p_{\rm 1}+\frac{1}{\eta_0}m_{\rm 2})\mathbf{x_0}+(p_{\rm 2}-\frac{1}{\eta_0}m_{\rm 1})\mathbf{y_0}\right],\vspace{.3cm}\\\displaystyle
\mathstrut_{\rm -z}\mathbf{E}_{\rm sc}=\gamma \hspace{1mm}
\left[ (p_{\rm 1}-\frac{1}{\eta_0}m_{\rm 2})\mathbf{x_0}+(p_{\rm 2}+\frac{1}{\eta_0} m_{\rm 1})\mathbf{y_0}\right],
\end{array}\l{eq:i}\f
where $\displaystyle \gamma=\frac{k^2}{4\pi \epsilon_0 r} e^{-j k r}$ is a parameter introduced for convenience.
Although these formulas are for the case of the incidence \r{eq:b}, similar formulas can be written also (with notations in bars) for the incidence defined in \r{eq:e}.
Combining equations \r{eq:i}, we find formulas for calculation of electric and magnetic dipole moments:
\e
\begin{array}{c}\displaystyle
p^{\pm}_{\rm 1}=\frac{1}{2\gamma}(\mathstrut_{\rm z} E^{\pm}_{\rm sc 1}+ \mathstrut_{\rm -z} E^{\pm}_{\rm sc 1})
,\qquad \displaystyle
p^{\pm}_{\rm 2}=\frac{1}{2\gamma}(\mathstrut_{\rm z} E^{\pm}_{\rm sc 2}+ \mathstrut_{\rm -z} E^{\pm}_{\rm sc 2})
,\vspace{.3cm}\\\displaystyle
m^{\pm}_{\rm 1}=\frac{\eta_0}{2\gamma}(\mathstrut_{\rm -z} E^{\pm}_{\rm sc 2}- \mathstrut_{\rm z} E^{\pm}_{\rm sc 2})
,\qquad\displaystyle
m^{\pm}_{\rm 2}=\frac{\eta_0}{2\gamma}(\mathstrut_{\rm z} E^{\pm}_{\rm sc 1}- \mathstrut_{\rm -z} E^{\pm}_{\rm sc 1})
.\end{array}\l{eq:j}\f
At this step, we are ready to write general formulas for calculating all the polarizability components in the $\_x_0\_y_0$-plane. 
First, we consider the case when the incident fields equal $\mathbf{E}_{\rm inc}=\eta_0 {H_0} \mathbf{x_0}$, $\mathbf{H}_{\rm inc}=\pm {H_0} \mathbf{y_0}$. Then, substituting \r{eq:j} in \r{eq:d}, we write the expressions for the polarizability components:
\e
\begin{array}{l}\displaystyle
\eexx=\frac{1}{4\gamma\E0} \left[
\mathstrut_{\rm z}  E^{+}_{\rm sc 1}+ \mathstrut_{\rm -z} E^{+}_{\rm sc 1}
+\mathstrut_{\rm z} E^{-}_{\rm sc 1}+ \mathstrut_{\rm -z} E^{-}_{\rm sc 1}\right],
\vspace{.3cm}\\\displaystyle
\emxy=\frac{1}{4\gamma\H0} \left[
\mathstrut_{\rm z} E^{+}_{\rm sc 1}+ \mathstrut_{\rm -z} E^{+}_{\rm sc 1}
-\mathstrut_{\rm z} E^{-}_{\rm sc 1}- \mathstrut_{\rm -z} E^{-}_{\rm sc 1}\right],
\vspace{.3cm}\\\displaystyle
\eeyx=\frac{1}{4\gamma\E0} \left[
\mathstrut_{\rm z} E^{+}_{\rm sc 2}+ \mathstrut_{\rm -z} E^{+}_{\rm sc 2}
+\mathstrut_{\rm z} E^{-}_{\rm sc 2}+ \mathstrut_{\rm -z} E^{-}_{\rm sc 2}\right],
\vspace{.3cm}\\\displaystyle
\emyy=\frac{1}{4\gamma\H0} \left[
\mathstrut_{\rm z} E^{+}_{\rm sc 2}+ \mathstrut_{\rm -z} E^{+}_{\rm sc 2}
-\mathstrut_{\rm z} E^{-}_{\rm sc 2}- \mathstrut_{\rm -z} E^{-}_{\rm sc 2}\right],
\vspace{.3cm}\\\displaystyle

\mexx=\frac{1}{4\gamma\H0} \left[
\mathstrut_{\rm -z} E^{+}_{\rm sc 2}- \mathstrut_{\rm z} E^{+}_{\rm sc 2}
+\mathstrut_{\rm -z} E^{-}_{\rm sc 2}- \mathstrut_{\rm z} E^{-}_{\rm sc 2}\right],
\vspace{.3cm}\\\displaystyle
\mmxy=\frac{\eta_0}{4\gamma\H0} \left[
\mathstrut_{\rm -z} E^{+}_{\rm sc 2}- \mathstrut_{\rm z} E^{+}_{\rm sc 2}
-\mathstrut_{\rm -z} E^{-}_{\rm sc 2}+ \mathstrut_{\rm z} E^{-}_{\rm sc 2}\right],
\vspace{.3cm}\\\displaystyle
\meyx=\frac{1}{4\gamma\H0} \left[
\mathstrut_{\rm z} E^{+}_{\rm sc 1}- \mathstrut_{\rm -z} E^{+}_{\rm sc 1}
+\mathstrut_{\rm z} E^{-}_{\rm sc 1}- \mathstrut_{\rm -z} E^{-}_{\rm sc 1}\right],
\vspace{.3cm}\\\displaystyle
\mmyy=\frac{\eta_0}{4\gamma\H0} \left[
\mathstrut_{\rm z} E^{+}_{\rm sc 1}- \mathstrut_{\rm -z} E^{+}_{\rm sc 1}
-\mathstrut_{\rm z} E^{-}_{\rm sc 1}+ \mathstrut_{\rm -z} E^{-}_{\rm sc 1}\right].
\end{array}\l{eq:k}\f
To clarify the notation here we can use an example. $\mathstrut_{\rm -z} E^{+}_{\rm sc 1}$ denotes the $\mathbf{x}_0$ projection of the scattered electric field in the $-\mathbf{z}_0$ direction if the scatterer is illuminated by the incident wave $\mathbf{E}_{\rm inc}=\eta_0 {H_0} \mathbf{x}_0$, $\mathbf{H}_{\rm inc}=+  {H_0} \mathbf{y}_0$.

Next, we study the case when the incident fields are defined as $\mathbf{E}_{\rm inc}=\eta_0{H_0} \mathbf{y}_0$, $\mathbf{H}_{\rm inc}=\pm {H_0} \mathbf{x}_0$. Likewise, the expressions for the other 8 polarizability components can be found:
\e
\begin{array}{c}\displaystyle
\eexy=\frac{1}{4\gamma\E0} \left[
\mathstrut_{\rm z} \bar{E}^{+}_{\rm sc 1}+ \mathstrut_{\rm -z} \bar{E}^{+}_{\rm sc 1}
+\mathstrut_{\rm z} \bar{E}^{-}_{\rm sc 1}+ \mathstrut_{\rm -z} \bar{E}^{-}_{\rm sc 1}\right],
\vspace{.3cm}\\\displaystyle
\emxx=\frac{1}{4\gamma\H0} \left[
\mathstrut_{\rm z} \bar{E}^{+}_{\rm sc 1}+ \mathstrut_{\rm -z} \bar{E}^{+}_{\rm sc 1}
-\mathstrut_{\rm z} \bar{E}^{-}_{\rm sc 1}- \mathstrut_{\rm -z} \bar{E}^{-}_{\rm sc 1}\right],
\vspace{.3cm}\\\displaystyle
\eeyy=\frac{1}{4\gamma\E0} \left[
\mathstrut_{\rm z} \bar{E}^{+}_{\rm sc 2}+ \mathstrut_{\rm -z} \bar{E}^{+}_{\rm sc 2}
+\mathstrut_{\rm z} \bar{E}^{-}_{\rm sc 2}+ \mathstrut_{\rm -z} \bar{E}^{-}_{\rm sc 2}\right],
\vspace{.3cm}\\\displaystyle
\emyx=\frac{1}{4\gamma\H0} \left[
\mathstrut_{\rm z} \bar{E}^{+}_{\rm sc 2}+ \mathstrut_{\rm -z} \bar{E}^{+}_{\rm sc 2}
-\mathstrut_{\rm z} \bar{E}^{-}_{\rm sc 2}- \mathstrut_{\rm -z} \bar{E}^{-}_{\rm sc 2}\right],
\vspace{.3cm}\\\displaystyle

\mexy=\frac{1}{4\gamma\H0} \left[
\mathstrut_{\rm -z} \bar{E}^{+}_{\rm sc 2}- \mathstrut_{\rm z} \bar{E}^{+}_{\rm sc 2}
+\mathstrut_{\rm -z} \bar{E}^{-}_{\rm sc 2}- \mathstrut_{\rm z} \bar{E}^{-}_{\rm sc 2}\right],
\vspace{.3cm}\\\displaystyle
\mmxx=\frac{\eta_0}{4\gamma\H0} \left[
\mathstrut_{\rm -z} \bar{E}^{+}_{\rm sc 2}- \mathstrut_{\rm z} \bar{E}^{+}_{\rm sc 2}
-\mathstrut_{\rm -z} \bar{E}^{-}_{\rm sc 2}+ \mathstrut_{\rm z} \bar{E}^{-}_{\rm sc 2}\right],
\vspace{.3cm}\\\displaystyle
\meyy=\frac{1}{4\gamma\H0} \left[
\mathstrut_{\rm z} \bar{E}^{+}_{\rm sc 1}- \mathstrut_{\rm -z} \bar{E}^{+}_{\rm sc 1}
+\mathstrut_{\rm z} \bar{E}^{-}_{\rm sc 1}- \mathstrut_{\rm -z} \bar{E}^{-}_{\rm sc 1}\right],
\vspace{.3cm}\\\displaystyle
\mmyx=\frac{\eta_0}{4\gamma\H0} \left[
\mathstrut_{\rm z} \bar{E}^{+}_{\rm sc 1}- \mathstrut_{\rm -z} \bar{E}^{+}_{\rm sc 1}
-\mathstrut_{\rm z} \bar{E}^{-}_{\rm sc 1}+ \mathstrut_{\rm -z} \bar{E}^{-}_{\rm sc 1}\right].
\end{array}\l{eq:l}\f
From \r{eq:k} and \r{eq:l} one can see that to extract one specific polarizability component of the particle by this method, we need to probe (or measure) the scattered fields only in two directions (at any arbitrary point in far-field). In order to find all 16 polarizability components, it is sufficient to know the scattered fields in two directions and to use only four different plane-wave illuminations.
In the next section, we show an example of implementation of this method for two different particles, one reciprocal and the other one non-reciprocal.

\section{Polarizability retrieval applied to reciprocal and non-reciprocal bi-anisotropic particles}\label{sec:3}

Here, we utilize the method for extracting polarizabilities of two bi-anisotropic particles that have been previously used as building blocks for metasurfaces possessing novel and exotic electromagnetic properties \cite{metamirrors,transparency}. In this paper we describe the method of extracting the polarizabilities in detail. The analyzed particles are electrically small, therefore, the present method can be applied. We determine the scattered fields by full-wave simulations using Ansoft High Frequency Structure Simulator. However, one can calculate the scattered fields using other approaches, e.g., based on the method of moments (MoM) or finite element method (FEM), or measure the far-fields experimentally.

\begin{figure}[H]
\centering
\begin{subfigure}{0.45\columnwidth}
  \centering
  \includegraphics[width=\columnwidth]{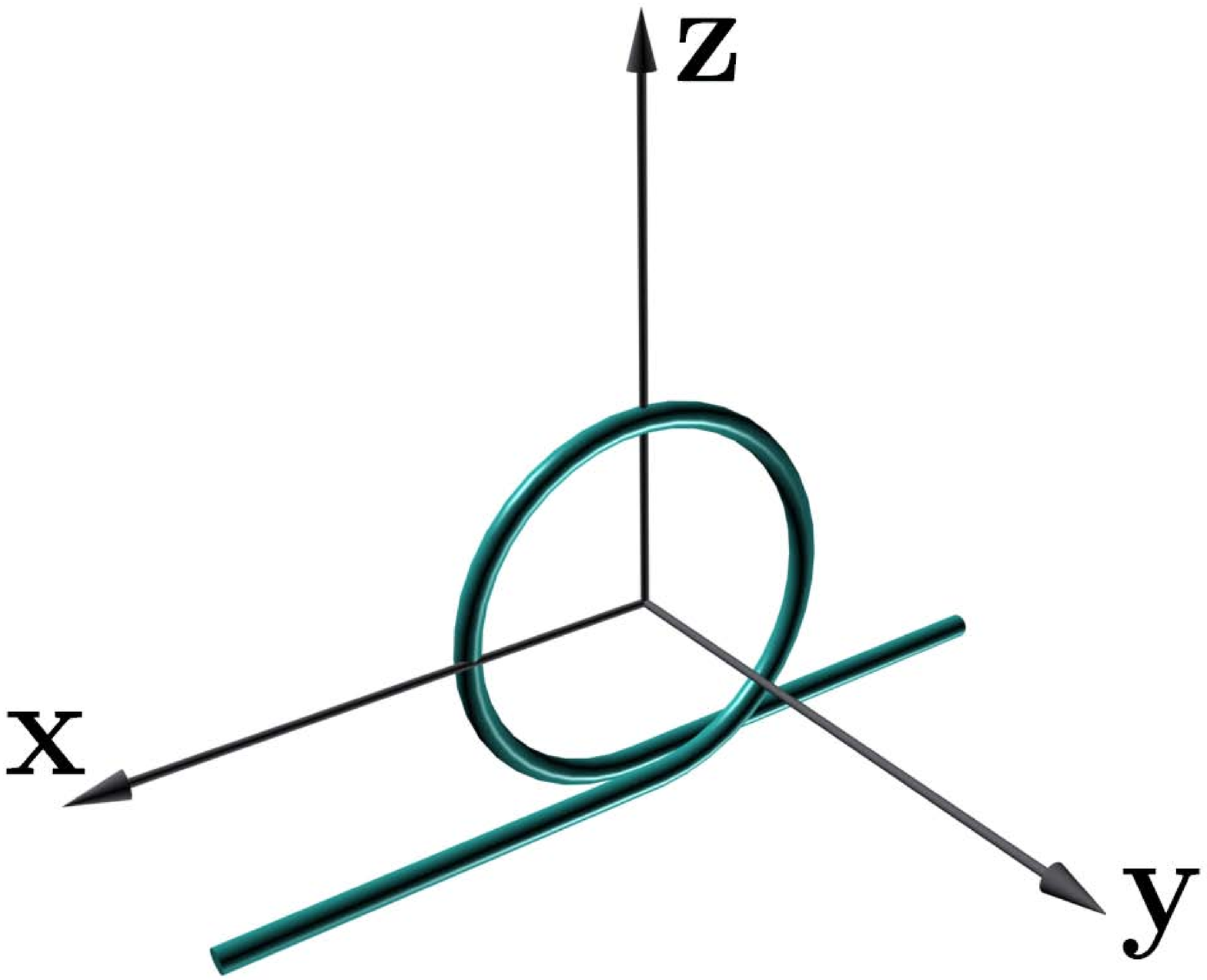}
  \caption{}
  \label{fig:omega}
\end{subfigure}
\begin{subfigure}{0.45\columnwidth}
  \centering
  \includegraphics[width=\columnwidth]{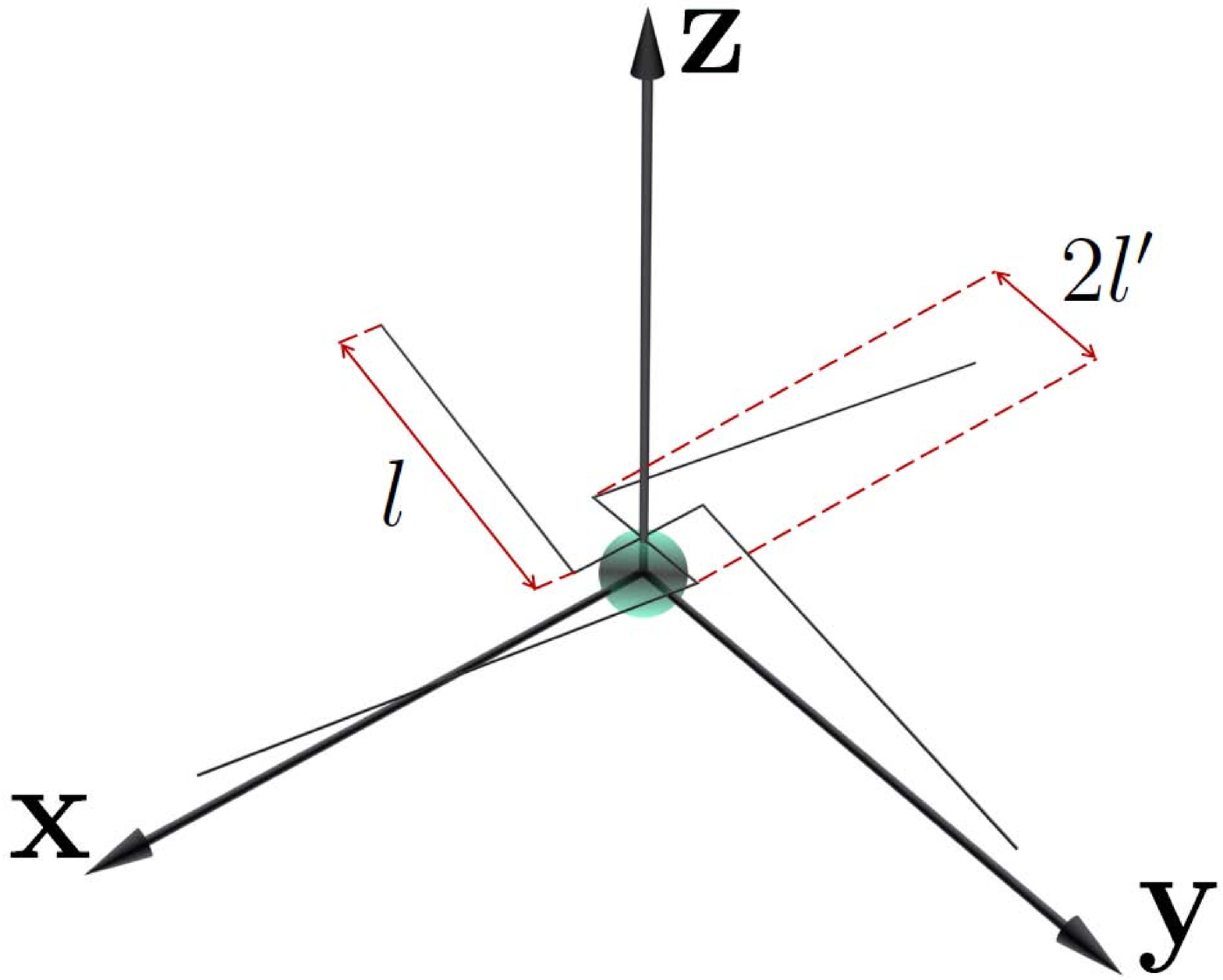}
  \caption{}
  \label{fig:moving}
\end{subfigure}
\caption{(a) The omega particle and the related coordinate system. (b) The non-reciprocal particle.}
\label{fig:test}
\end{figure}

\begin{figure}[H]
\centering
\epsfig{file=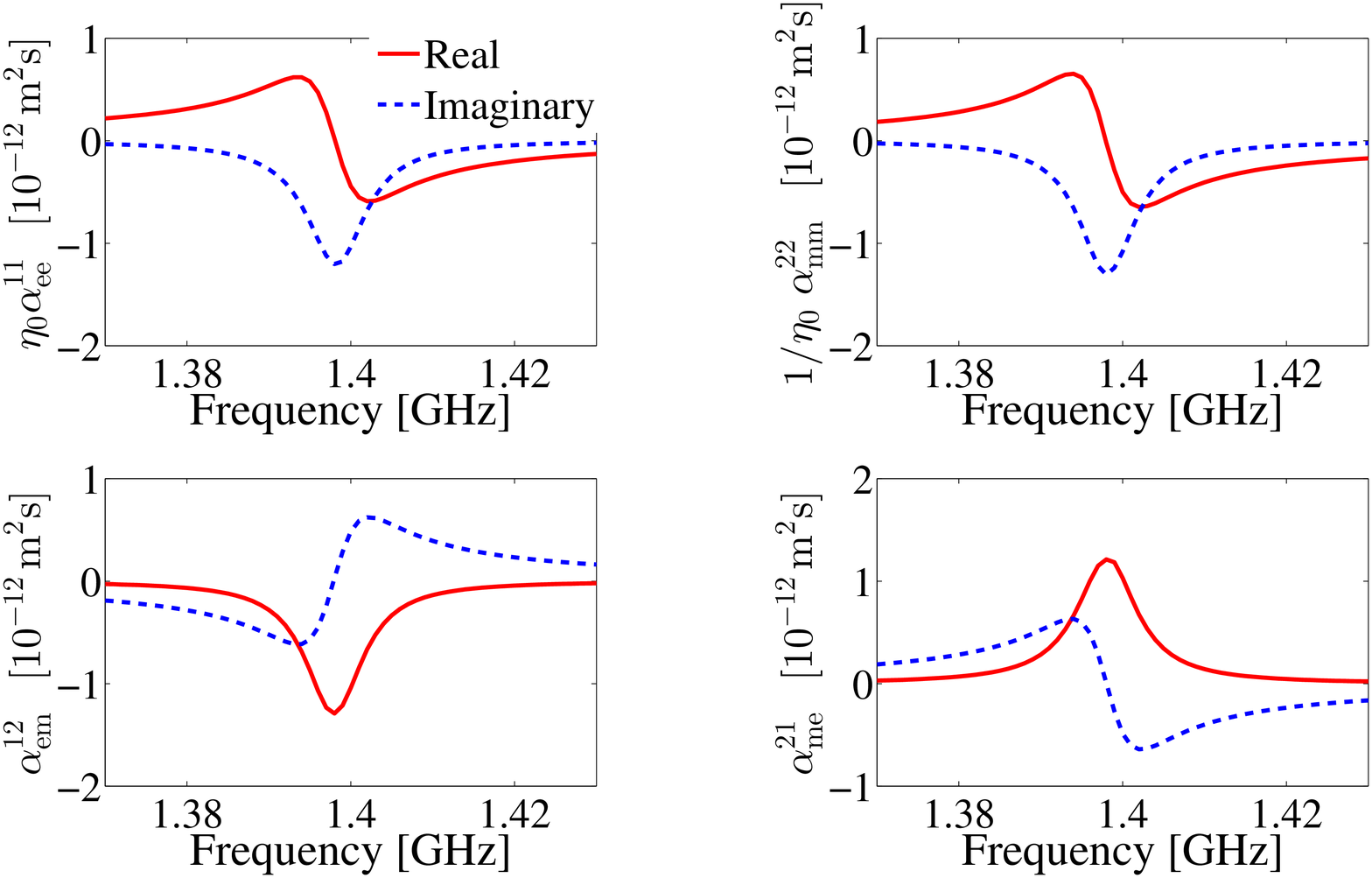, width=0.75\textwidth}
\caption{Polarizabilities of the omega particle, normalized to the free-space impedance.}
\label{fig:omega2}
\end{figure}

The first example, which we consider here, is a reciprocal omega particle \cite{serdukov} shown in Fig.~\ref{fig:omega}.
It was previously shown that this kind of inclusions can be utilized in thin composite metamirrors (full-reflection layers) which allow full control over phase of reflection \cite{metamirrors}.
The dimensions of the particle under study are as follows: The radius of the loop is \mbox{$r=7.45$ mm},
the half-length of the electric dipole is \mbox{$d=18.1$ mm},
the radius of the wire is \mbox{$r_0=0.5$ mm}, and the pitch is $1.45$~mm. The material of the particle is PEC. In the defined coordinate system only four polarizability components of the particle are significant \cite{serdukov}: $\eexx$, $\mmyy$, $\emxy$, and $\meyx$. As it is seen from \r{eq:d}, we can find all these components using two incident waves with the polarization states defined by \r{eq:b}. In order to determine the electric and magnetic dipolar moments in \r{eq:d}, we probe the fields scattered by the particle in the $+\mathbf{z_0}$ and $-\mathbf{z_0}$ directions, as it is dictated by \r{eq:j}. Next, using the final formulas \r{eq:k}, we plot the polarizability components of the particle versus frequency (see Fig.~\ref{fig:omega2}).
As it is seen, at the resonance frequency of the particle electric and magnetic polarizabilities become purely imaginary while the electromagnetic polarizabilities are real. It was expected, since all the reactances are compensated and only dissipative terms remain for the particle at resonance. The condition $\emxy=-\meyx$ holds for the particle, as it must be in accordance with the Onsager-Casimir principle \cite{Landau}. Also it can be seen from Fig.~\ref{fig:omega2} that the electric and magnetic polarizabilities satisfy the balance condition $\n_0\eexx=\mmyy/\n_0$. This corresponds to the case of extreme response of balanced bi-anisotropic particles \cite{balanced}, and it was the design requirement in \cite{metamirrors}.

\begin{figure}[H]
\centering
\epsfig{file=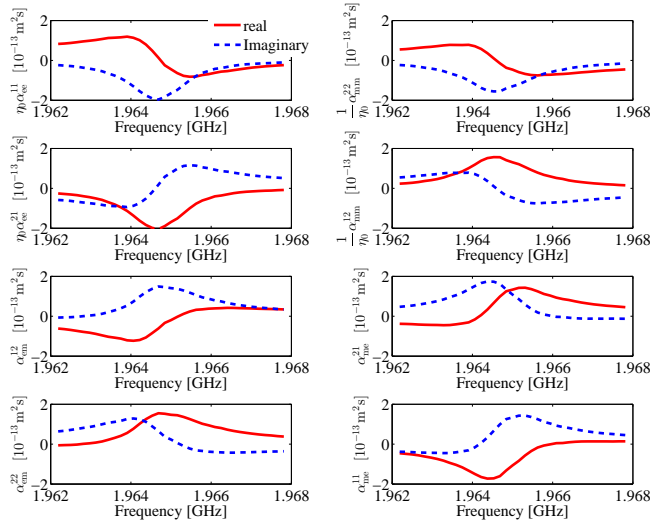, width=0.85\textwidth}
\caption{Polarizabilities of the non-reciprocal particle.}
\label{fig:moving2}
\end{figure}

As another example, we analyze a non-reciprocal particle possessing moving and chiral electromagnetic couplings (see Fig.~\ref{fig:moving}). A planar array of these particles acts as a non-reciprocal one-way transparent ultimately thin layer \cite{transparency}. The layer is transparent from one side while from the opposite side the layer acts as a twist-polarizer in transmission. A ferrite sphere magnetized by external bias field is the non-reciprocal element in the particle. The ferrite sphere with the radius \mbox{$a=1.65$ mm} is coupled to metal elements with the dimensions \mbox{$l=18$ mm} and \mbox{$l'=3$ mm}. The radius of the copper wire is \mbox{$\delta=0.05$ mm}. The ferrite material is yttrium iron garnet: The relative permittivity $\epsilon_r=15$, the dielectric loss tangent $\tan\delta=10^{-4}$, the saturation magnetization \mbox{$M_S=1780$ G}, and the full resonance linewidth \mbox{$\Delta H=0.2$ Oe}. The $+\mathbf{z}_0$ internal bias field is \mbox{$H_b=9626$ A/m}, corresponding to the desired resonance frequency. In the defined coordinate system, significant polarizabilities of the particle are those in the $\_x_0\_y_0$-plane. In the same way as for the omega particle, we find the normalized polarizabilities for moving-chiral particle (see Fig.~\ref{fig:moving2}).
One can see from Fig.~\ref{fig:moving2} that the Onsager symmetry relations $\emyy=-\mexx$ and $\emxy=\meyx$ hold for the particle.

\section{Conclusions}
Here, we have presented a method which allows us to find all polarizability tensor components for an electrically small arbitrary bi-anisotropic particle with any complex shape and internal structure. We have assumed that only dipolar moments are significant in the particle, that is, that the particle is electrically small. In comparison to other known methods, this method requires less complicated calculations. To determine one specific polarizability component of the particle, the method requires probing of the particle response only by plane waves in two directions. For determining all the 36 tensor components it is sufficient to find the scattered fields only in 6 directions for 12 different incidences. The scattering response is measured only in the far zone and only in a few directions (6 for the most general particle). Due to simplicity of the method, the method can be utilized also experimentally.
In order to demonstrate and illustrate the concept, we have derived formulas for 16 polarizability components. One can similarly derive the formulas for the other 20 components.
In the paper, the polarizability retrieval method has been applied for two specific bi-anisotropic particles, to give particular examples. The method can be used to determine and optimize single inclusions in metamaterials and metasurfaces with the goal to achieve desired electromagnetic properties of the whole structure.


\section*{References}
\begin{thebibliography}{10}
\bibitem{Lewin}
L.~Lewin, The electrical constants of a material loaded with spherical
particles, Radio and Communication Engineering 94 (1947) 65--68.

\bibitem{Sihvola}
A.~Sihvola, Electromagnetic Mixing Formulas and Applications, 1st ed., London, IEEE Publishing, 1999.

\bibitem{analytical}
S.~A.~Tretyakov, Analytical Modeling in Applied Electromagnetics, Norwood, Artech House Publishers, 2003.


\bibitem{Holloway}
E.~F.~Kuester, M.~A.~Mohamed, C.~L.~Holloway, Averaged transition conditions for electromagnetic field at a metafilm, IEEE Transactions on Antennas Propagation 51 (2003) 2641-–2651.


\bibitem{Alu_ho}
Y.~Zhao, N.~Engheta, A.~Al\`u, Homogenization of plasmonic metasurfaces modeled as transmission-line loads, Metamaterials 5(2011) 90–-96.



\bibitem{Teemu}
T.~Niemi, A.~Karilainen, S.~Tretyakov, Synthesis of polarization transformers, IEEE Transactions on Antennas Propagation 61 (2013) 3102--3111.


\bibitem{Mie}
G.~Mie, Beitr\"age zur optik tr\"uber medien speziell kolloidaler
metall\"osungen, Annalen der Physik 330 (1908) 377--445.

\bibitem{Lorenx}
L.~V.~Lorenz, Sur la lumi\`ere r\'efl\'echie et r\'efract\'ee par une
sph\`ere transparente, Oeuvres Scientifiques, Librairie Lehman et Stage,
Copenhagen (1898) 405--529.

\bibitem{Bohren}
C.~F.~Bohren, D.~R.~Huffman, Absoption and Scattering of Light by Small Particles, 1st ed., New York, Wiley, 1983.

\bibitem{Waterman}
P.~C.~Waterman, Symmetry, unitary, and geometry in electromagnetic
scattering, Physical Review D 3 (1971) 825--839.

\bibitem{Menzel}
S.~M\"uhlig, C.~Menzel, C.~Rockstuhl, F.~Lederer, Multipole analysis of meta-atoms, Metamaterials 5 (2010) 64--73.

\bibitem{Arango}
F.~B.~Arango, A.~F.~Koenderink, Polarizability tensor retrieval for magnetic and plasmonic antenna design, New Journal of Physics 15 (2013) 073023.

\bibitem{Fanyaev}%
V.~S.~Asadchy, I.~A.~Faniayeu, Simulation of the electromagnetic properties of one-turn and double-turn helices with optimal shape, which provides radiation of a circularly polarized wave, Journal of Advanced Research in Physics 2 (2011) 011107.

\bibitem{antenna}
S.~A.~Tretyakov, F.~Mariotte, C.~R.~Simovski, T.~G.~Kharina, J.-P.~Heliot, Analytical antenna model for chiral scatterers: Comparison with numerical and experimental data, IEEE Transactions on Antennas and Propagation 44 (1996) 1006--1014.


\bibitem{serdukov}%
A.~N.~Serdyukov, I.~V.~Semchenko, S.~A.~Tretyakov, A.~Sihvola, Electromagnetics of Bi-Anisotropic Materials: Theory and Applications, Amsterdam, Gordon and Breach Science Publishers, 2001.

\bibitem{tretyakov32}
S.~A.~Tretyakov, A.~H.~Sihvola, A.~A.~Sochava, C.~R.~Simovski, Magnetoelectric interactions in bi-anisotropic media, Journal of Electromagnetic Waves and Applications 12 (1998) \mbox{481--497}.

\bibitem{Definition}
J.~D.~Jackson, Radiating Systems, Multipole Fields and Radiation, 3rd ed., New York, Wiley (1999) 407--455.

\bibitem{metamirrors}
Y.~Ra'di, V.~S.~Asadchy, S.~A.~Tretyakov, Tailoring reflections from thin composite metamirrors (2013) arXiv:1401.1677.

\bibitem{transparency}
Y.~Ra'di, V.~S.~Asadchy, S.~A.~Tretyakov, One-way transparent sheets (2013) arXiv:1310.4586.

\bibitem{Landau}
L.~D.~Landau, E.~M.~Lifshitz, Electrodynamics of Continuous
Media, 2nd ed., Oxford, England, Pergamon Press, 1984.

\bibitem{balanced}
Y.~Ra'di, S.~A.~Tretyakov, Balanced and optimal bi-anisotropic particles: Maximizing power extracted from electromagnetic fields, New Journal of Physics 15 (2013) 053008.


\end{thebibliography}

\end{document}